# Antimony doped Tin Oxide/Polyethylenimine Electron Selective Contact for reliable and light soaking-free high Performance Inverted Organic Solar Cells

*Efthymios Georgiou[1], Ioannis T. Papadas[1], Ioanna Antoniou, [1]Marek F. Oszajca[2], Benjamin Hartmeier[2], Michael Rossier[2], Norman A. Luechinger[2] and Stelios. A. Choulis[\*,1]*

[1] Molecular Electronics and Photonics Research Unit, Department of Mechanical Engineering and Materials Science and Engineering, Cyprus University of Technology, 45 Kitiou Kyprianou Street, Limassol, 3603, Cyprus

[2] Avantama Ltd, Staefa, Laubisrutistr. 50, CH-8712, Switzerland

**Keywords:** metal oxides, antimony doped tin oxide, polyethylenimine, interfaces, electron selective contact, electrodes, light-soaking, organic photovoltaics

\* E-mail: stelios.choulis@cut.ac.cy

## Abstract

We have demonstrated a high-performance low temperature solution processed electron selective contact consisting of 10 at% antimony doped tin oxide (ATO) and the neutral polymer polyethylenimine (PEI). Inverted organic photovoltaics (OPVs) utilizing ATO/PEI as electron selective contact exhibited high power conversion efficiencies for both the reference P3HT: PCBM and the non-fullerene based P3HT- IDTBR active layer OPV material systems. Importantly it is shown that the proposed ATO/PEI carrier selective contact provides light soaking-free inverted OPVs. Furthermore, by increasing the thickness of ATO layer from 40 to 120 nm the power conversion efficiency of the corresponding inverted OPVs remain unaffected a parameter which indicates the potential of the proposed ATO/PEI carrier selective contact for high performance light-soaking-free and reliable roll-to-roll printing solutions processed inverted OPVs.

# Introduction

The ability to fabricate ultrathin, lightweight, flexible and low-cost photovoltaics trigger the research community and industries in a depth investigation of organic photovoltaics (OPVs) during the last two decades. The power conversion efficiency (PCE) and the lifetime performance of OPVs are continuously improving. Recent progress in OPVs overcame the barrier of 10% certified power conversion efficiency.[1] In addition, the progress of the lifetime performance of OPVs during the last years resulted in OPVs with promising long term stability.[2] Development of high performance solution processed electronic materials is one of the key OPV product development targets since the application of roll to roll printing manufacturing determine the potential of OPVs for future low-cost next generation photovoltaics.

The development of inverted OPVs provided an interesting architecture [3], since it allowed studying fundamental processes in organic bulk heterojunction architecture, including the vertical phase segregation of polymer / fullerene blends as well as the charge selectivity of the metal-oxide based carrier selective contacts.[3][4] Metal oxides have been extensively used as n-type or p-type buffer layers in inverted OPVs due to their high transparency, relatively high conductivity, tunable work function and solution processability. Furthermore, the advantage of solution processing of these materials at low temperatures provide opportunities for low cost, reliable and roll-to-roll process up-scalable OPVs. Solution processed doped metal oxides with high transparency and conductivities could meet the requirements for reliable and up-scalable roll-to-roll printing process inverted OPVs since thicker buffer layers (100-200 nm) can be deposited without affecting significantly the device performance.[5,6] $SnO_2$ is an n-type, wide bandgap semiconductor that, when doped, can exhibit good transparency throughout the visible range of the solar spectrum and low electrical resistivity. Doped tin oxides are commonly used as buffer layers for a wide range of applications including solar cells.[7][8]

The first reported metal oxide interface modification for improving the performance of inverted OPVs included the incorporation of a polyoxyethylene tridecyl ether (PTE) interfacial layer between ITO and solution-processed titanium oxide [4]. More recently, Poly[(9,9-bis(3′ - (N,N-dimethylamion)propyl)-2,7-fluorene)-alt-2,7-(9,9-dioctyl)-fluorene (PFN) or Polyethylenimine (PEI) interface modification of metal-oxide based materials have been reported in the literature by several methods such as dipping, bi-layer or as additive in the precursor and have been resulted to improved OPVs device performance.[9–13]. In more details, it has been shown that incorporation of PEI improves the morphology and structural order of ZnO and thereby the electron mobility in the vertical direction. Most importantly PEI interface modification is widely used to reduce the energy levels of metal oxides and consequently the work function of bottom electrodes of inverted OPVs and thereby assist in suitable band alignment of the metal oxide with the OPV based active layer.[14] The work function modification is caused by the formation of an interfacial dipole at the interface which induces a vacuum level shifting and thereby a work function reduction.[15][16][17] As a result, such interface modification enhances the device performance of OPVs mainly due to improved FF and Voc values.

In this work we present a highly conductive and transparent 10 at% antimony doped tin oxide (ATO) modified with ultrathin PEI  interfacial layer as a high performance, reliable and light soaking-free electron selective contact for inverted OPVs. Since the PTE interfacial layer in the range of 5 nm, the PEI processing step can be also considered as interface conditioning step of ATO. By applying the proposed ATO/PEI electron selective contact inverted OPV PCEs of 3.8% were achieved with the reference P3HT: PCBM active layer system. The current density characteristics of the corresponding devices remain unaffected from UV light soaking, which is an important parameter for newly developed metal-oxide based carrier selective contacts. Furthermore, increasing the thickness of ATO from 40 nm to 120 nm the power conversion efficiency of the corresponding ATO/PEI electron selective contact based inverted OPVs remain unaffected a parameter which indicates the potential of ATO/PEI carrier selective contact for reliable roll-to-roll printing processed OPVs. Importantly, the proposed ATO/PEI electron selective contact was also applied to inverted OPVs incorporating non-fullerene IDTBR acceptors. The corresponding non-fullerene based inverted OPVs using ATO/PEI electron selective contact obtain PCE of 5.74% revealing the suitability of ATO/PEI electron selective contact for highly efficient non-fullerene acceptor based inverted OPVs.

## Antimony Doped Tin Oxide Synthesis and Characterization

The 10 at% antimony doped tin oxide (10 at% Sb: $SnO_2$) solution in mixture of butanols (ATO) was developed from Avantama. The ATO powders were synthesized from Avantama Ltd by flame spray pyrolysis. For flame spray pyrolysis a metal organic precursor was prepared by dissolving a tin salt and an antimony salt in alkyl carboxylic acid and toluene. The metal organic precursor was fed (9 mL/min) to a spray nozzle, dispersed by oxygen (14 L/min) and ignited by a premixed methane–oxygen fame ($CH_4$: 1.2 L/min, $O_2$: 2.2 L/min). The off-gas was filtered through a metal filter (BOPP Ag) by a vacuum pump (Busch, Mink MM 1142 BV) at about 20 $m^3$/h. The obtained nano-powder was collected from the metal filter. The as-prepared ATO nanoparticles were stabilized in mixture of butanols at 2.5 wt% using an undisclosed Avantama Ltd surfactant.

Figure 1a shows the X-ray diffraction pattern for the tin oxide with 10 at% doped antimony. All the diffraction pattern shows characteristic tin oxide peaks with tetragonal structure (Cassiterite, Space group: $P4_2$/mnm, unit cell: $a$ = 4.738 Å and $c$ = 3.187 Å, PDF #41-1445).

The average domain size of $SnO_2$ crystallites calculated from Scherrer analysis of the (110) peak width is ~5 nm. It seems that Sb dopant does not form a second phase either in or with the $SnO_2$.

Figure 1b, shows the corresponding Tauc plot obtained by the optical absorption spectrum of the ATO films fabricated on a quartz substrate for direct allowed transition [$(\alpha E)^2$ *versus* photon energy (E)]. The ATO films exhibit an intense optical absorption onset in the UV region, which is associated with an optical band gap of 3.9 eV (Fig. 1b). The observed wide optical band gap implies a high optical transparency of the proposed ATO buffer layers in the desired wavelength range. Figure 1c shows the transparency of ATO buffer layers at different thickness. As shown in Figure 1c ATO buffer layers present high transparency in the visible spectrum and one can observe a slightly reduction of transmittance while the thickness of ATO layer increases to 240 nm. The excellent transparency of the proposed inverted OPV bottom electrode that is based on indium doped tin oxide (ITO)/ATO/PEI is shown in Figure 1d, which also includes the transmittance measurements of ITO, ITO/PEI bottom electrodes. In addition, the conductivity of the ATO based buffer layers was measured with four-point probe technique resulting in values in the order of $10^{-3}$ S.cm$^{-1}$, which is a targeted

conductivity value of doped oxides for OPVs application.

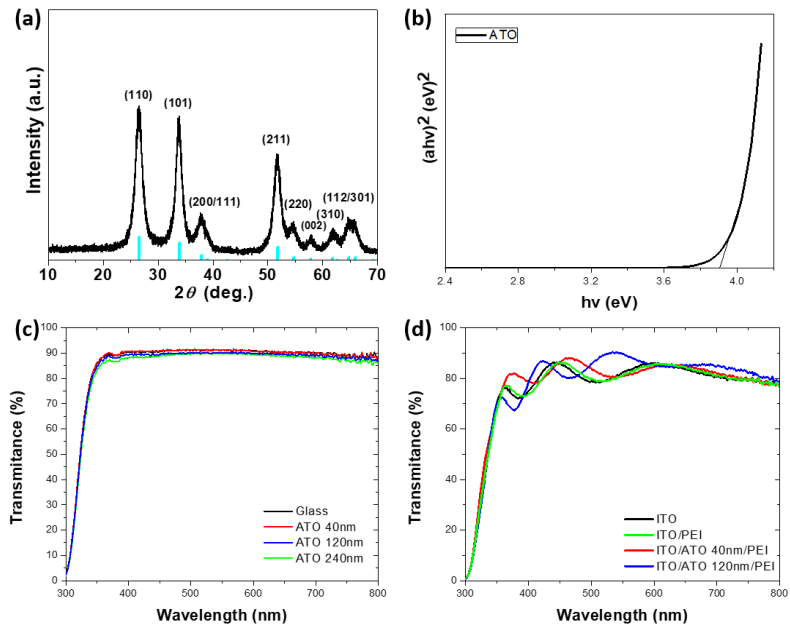

**Figure 1: (a) XRD pattern of ATO nanoparticles. The standard pattern of SnO₂ (PDF #41-1445) is also given for comparison (cyan line). (b) Tauc's plot, (α.E)² vs. photon energy (E), showing an energy band gap 3.9 eV of ATO NPs film fabricated on quartz substrate. (c) Transmittance of ATO at different layer thicknesses,.(d) Transmittance of bottom electrode: ITO, ITO/PEI, ITO/ATO 40 nm/PEI, ITO 120 nm PEI.**

To investigate the surface topography of ITO and ATO before and after the deposition of the ultrathin PEI interfacial layer, atomic force microscopy (AFM) images were obtained and illustrated in Figure 2.

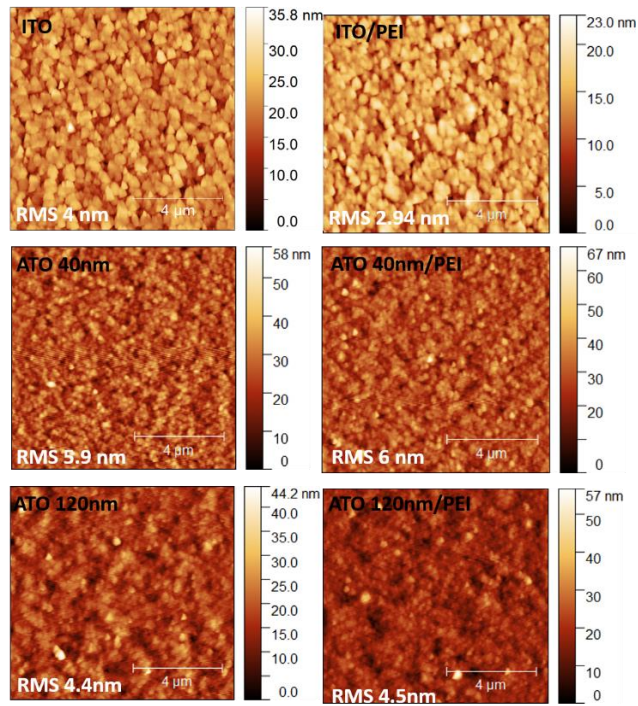

**Figure 2: AFM images of ITO, ITO/PEI, ATO 40 nm, ATO 40 nm /PEI, ATO 120 nm and ATO 120 nm/ PEI.**

As can be seen from Figure 2 pristine ITO presents root mean square roughness (Rq) in the range of 4 nm and is reduced at 2.94 nm after the deposition of PEI. In general, ATO layers present smooth surface topography and relatively low Rq. The 40 nm ATO and 120 nm ATO layers present relatively the same Rq before and after PEI interface modification, from 5.9 nm to 6 and from 4.4 nm to 4.5 nm, respectively.

Figure 3a and 3b shows the illuminated and dark current density versus voltage (JV) characteristics of inverted P3HT:PCBM based OPVs with different carrier selective contact(CSC) using the following device configuration: ITO/CSC/P3HT:PCBM/MoO$_3$/Ag.

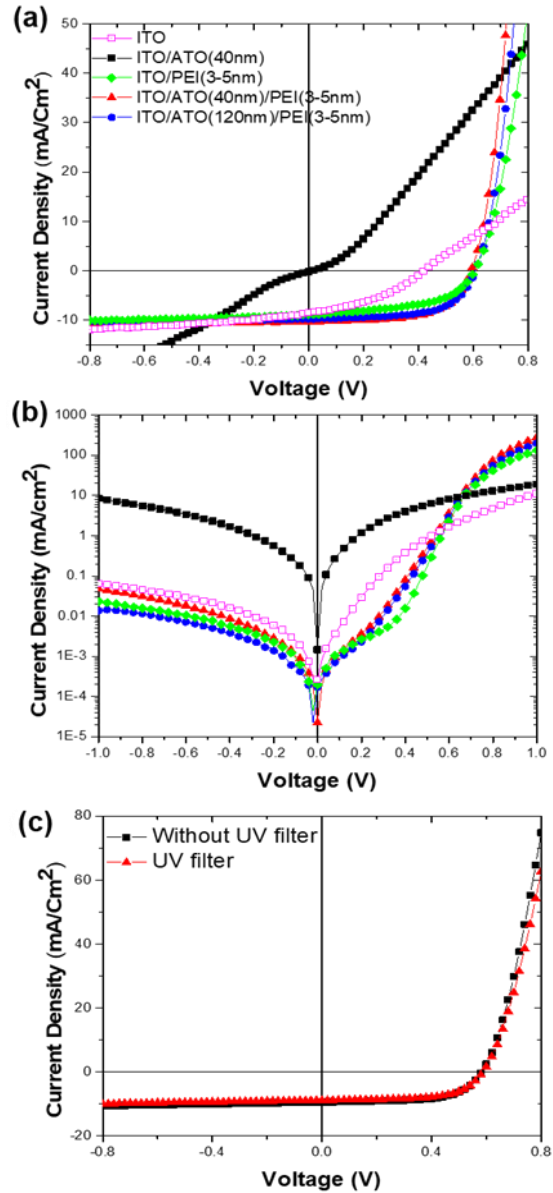

**Figure 3: (a) Illuminated JV characteristics of inverted OPVs with: ATO (black squares), ATO (40 nm)/PEI (red rectangles), ATO (120 nm)/PEI (blue circles), PEI (green diamonds) and ITO only (magenta squares). (b) Dark JV characteristics. (c) JV characteristics of inverted OPVs with ATO (40 nm)/PEI as electron selective contact under AM1.5 illumination without and with UV blocking filter (λ > 400 nm).**

As shown in Figure 3a and 3b OPVs using just ITO have shown high leakage current and high series resistance, Rs, showing poor selectivity and thus low PCE of 1.4%. On the other hand, OPVs with ITO/PEI (modification of ITO with PEI) have shown an improvement in carrier selectivity which is reflected to better FF value and thus better device performance with PCE of 2.9% as depicted in Table 1 which summarize the corresponding OPV device performance parameters. The devices using only (40 or 120 nm) thick 10 at% ATO as buffer layer function as a resistor (Figure 3) showing that an energetic barrier is created at the bottom electrode. For the above reason the device performance parameters for inverted OPVs using ITO/10 at% doped ATO bottom electrode within the table 1 are indicated as non-functional (NF). Importantly, by the using the PEI interface modification inverted OPVs using ITO/ATO/PEI as bottom electrode were performed efficiently. The proposed ATO/PEI carrier selective contact significantly improved electron carrier selectivity in inverted OPVs. Inverted P3HT:PCBM based OPVs with 0.6 V open circuit voltage (Voc), 10.3 mA/cm$^2$ short circuit current (Jsc) and 61% FF and 3.8% PCE were achieved by using ITO/ATO/PEI bottom electrode as illustrated in Figure 3. According to the literature the LUMO level of PC[60]BM is -4.7 V which is a value that doesn't match with the energy levels of the 10 at% doped ATO.[18] As previously reported in the literature PEI based interface modification cause reduction of the metal-oxide work function.[9] The high photovoltaics performance of P3HT:PCBM based inverted OPV reported in this paper by using ITO/ATO/PEI bottom electrodes indicates suitable energy levels matching between the ATO/PEI and PC[60]BM that resulted to efficient electron carrier selectivity for the inverted OPVs under investigation.

Furthermore, due to high conductivity and transparency of ATO by increasing the thickness of doped 10 at% doped ATO at 120 nm the performance of P3HT:PCBM based inverted OPVs using ITO/120 nm ATO/PEI bottom electrode remains unaffected (3.7% PCE). These results indicating that the PEI interface modification does not affect conductivity of the 10 at% doped ATO in the vertical direction and therefore the proposed thick ATO/PEI electron selective contact can provide a reliable buffer layer for the roll-to-roll printing processing of solution processed photovoltaics.

**Table 1: JV device performance parameters of inverted P3HT: PCBM OPVs with electron selective contact (ESC): ATO (40nm), ATO (40nm)/PEI, ATO (120nm)/PEI, PEI and without ETL (ITO only).**

| OPVs with ESC: | $V_{oc}$ (V) | $J_{sc}$ (mA.cm$^{-2}$) | FF (%) | PCE (%) |
|---|---|---|---|---|
| ITO | 0.42 | 8.5 | 39 | 1.4 |
| ITO/PEI | 0.60 | 8.6 | 56 | 2.9 |
| ITO/ATO 40 nm & 120 nm | NF | NF | NF | -NF |
| ITO/ATO 40 nm/PEI | 0.60 | 10.3 | 61 | 3.8 |
| ITO/ATO 120 nm/PEI | 0.62 | 9.8 | 61 | 3.7 |

A obstacle of inverted OPVs using metal oxides as electron transporting layer (ETL) is the light soaking effect, in which some metal-oxide buffer layers were reported to need UV light in order to be activated.[19]In the absence of UV light (light-soaking) OPVs that use metal oxides such as TiOx or ZnO in some cases were reported to present S-shape JV characteristics and poor electron selectivity resulting in low FF factor and PCE values [4, 14]. Solar spectrum includes UV radiation, and is well known that UV light is a factor that negatively affecting OPVs stability. To improve OPV lifetime encapsulation (packaging) of OPVs usually incorporate UV filter. Taking into consideration that an organic solar cell product will need the presence of a UV-filter as part of the packaging to prevent photo-oxidisation of the organic based active layer, light-soaking free carrier selective contacts are essential for the product development of inverted OPVs. It has been shown that in some cases this phenomenon can be mitigated with the use of doped metal-oxides.[5] In addition, SnOx was reported as light soaking free ETL.[20][21] Figure 3c shows the  illuminated JV characteristics of inverted P3HT:PCBM based OPVs using ATO/PEI electron selective contact  with and without UV-filter. To investigate light soaking issues illuminated JV characteristics of inverted P3HT:PCBM based OPVs with ATO/PEI were obtained with UV filter (λ>400 nm) showing similar performance with the illuminated JV characteristics obtained without UV filter as shown in Figure 3c. These results provide a strong indication that the performance of the proposed ATO/PEI electron selective contact for inverted OPVs is not affected from UV-light and thus can provide light soaking free inverted structured OPVs.

Novel non-fullerene acceptor materials such as ITIC and IDTBR have already been used with newly synthesized conjugated polymeric donors such as PTB7-th and PBDBT achieving efficiencies that exceed 10%.[22] This remarkable progress in terms of device efficiency demonstrates the potential of large-scale production of non-fullerene based OPVs. When mixing the well-known P3HT with the non-fullerene acceptor IDTBR (LUMO: -3.9 eV, HOMO: -5.45 eV), OPVs with 6.4% PCE can be achieved.[23][24] Therefore, to further investigate the functionality of the proposed ATO/PEI electron selective contact , inverted OPVs with the well-known polymer donor P3HT and the non-fullerene acceptor IDTBR were fabricated and their JV characteristics under dark and light are shown in Figure 4. While, table 2 summarizes the performance parameters of the non-fullerene (P3HT: IDTBR) based inverted OPVs using the ITO/ATO/PEI bottom electrode.

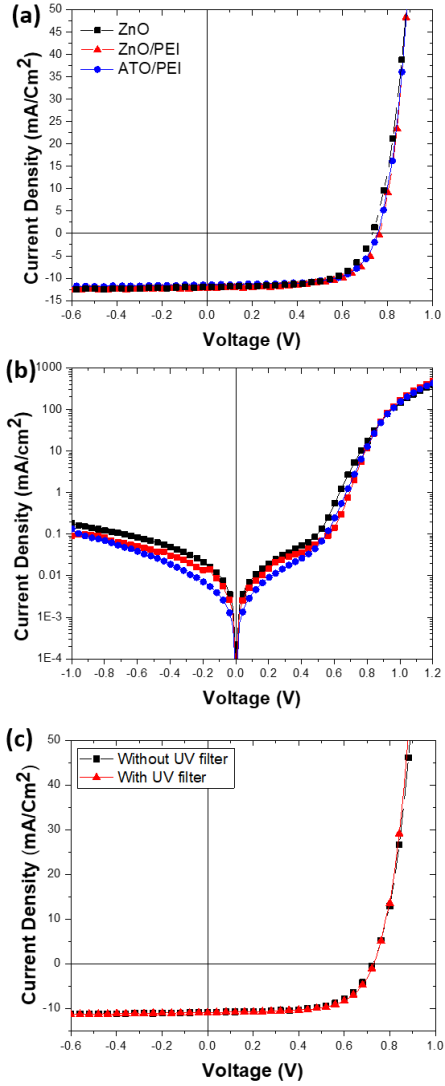

**Figure 4: (a) Illuminated JV characteristics of inverted OPVs with P3HT: IDTBR as active layer and ETL: ZnO (black squares), ZnO/PEI (red rectangles), ATO/PEI (blue circles) (b) Dark JV characteristics, (c) JV characteristics of inverted OPVs with ATO (40 nm)/PEI as electron selective contact under AM1.5 illumination without and with UV blocking filter (λ > 400 nm).**

**Table 2: JV device performance parameters of inverted OPVs with P3HT: IDTBR as active layer and ESC: ZnO, ZnO/PEI, ATO/PEI**

| OPVs with ETL: | $V_{oc}$ (V) | $J_{sc}$ (mA.cm$^{-2}$) | FF (%) | PCE (%) |
|---|---|---|---|---|
| **ITO/ZnO** | 0.74 | 12.02 | 62.9 | 5.61 |
| **ITO/ZnO/PEI** | 0.76 | 12.22 | 63.6 | 5.93 |
| **ITO/ATO/PEI** | 0.76 | 11.46 | 65.7 | 5.74 |

PCE of 5.74% was achieved for the proposed ATO/PEI electron selective contact whereas the reference inverted OPVs with ZnO and ZnO/PEI shown PCE of 5.61% and 5.93% respectively. These results importantly shown that the proposed ATO/PEI electron selective contact can be effectively applied to highly efficient non-fullerene acceptor inverted structured based OPVs. To further confirm the light soaking free inverted OPV device performance for the proposed ATO/PEI carrier selective contact illuminated JV characteristics of non-fullerene based P3HT: IDTBR inverted OPVs with ATO/PEI electron selective contact were obtained with UV filter ($\lambda$>400 nm) showing similar performance with the illuminated JV characteristics obtained without UV filter as shown in Figure 4c.

To conclude, a fundamental understanding of metal oxide interfaces and their manipulation is expected to be crucial to the continued progress of solution processed OPVs. We have demonstrated that the interface of the highly conductive and transparent solution processed 10 at% doped ATO can be effectively modified by the neutral polymer (PEI). Inverted fullerene and non-fullerene acceptor based OPVs utilizing ATO/PEI as electron selective contact exhibited optimized and light soaking-free PCE for both P3HT: PCBM (3.8 %, PCE) and P3HT: IDTBR (5.7 %, PCE) active layer material systems. The high transparency and conductivity of 10 at% doped ATO based buffer layers allowed the solution processing of high performance inverted OPVs utilizing thick ATO/PEI carrier selective contact. An important parameter for the reliable roll-to-roll printing process of OPVs and other printed opto-electronic applications.

**Supplementary information:** See supplementary material for details of the proposed electron transporting layer, processing of the materials, and OPV device fabrication. Additional information of material characterization, surface topography, XRD and other experimental results are included.

**Acknowledgements:** This project received funding from the European Research Council (ERC) under the European Union's Horizon 2020 research and innovation program (Grant Agreement No. 647311).

# Supplementary Information

## Antimony doped Tin Oxide/Polyethylenimine Electron Selective Contact for reliable and light soaking-free  high Performance Inverted Organic Solar Cells

*Efthymios Georgiou[1], Ioannis T. Papadas[1], Antoniou I. [1]* Marek F. Oszajca[2], Benjamin Hartmeier[2], Michael Rossier[2], *Norman A. Luechinger[2] and Stelios. A. Choulis[*,1]*

[1] Molecular Electronics and Photonics Research Unit, Department of Mechanical Engineering and Materials Science and Engineering, Cyprus University of Technology, 45 Kitiou Kyprianou Street, Limassol, 3603, Cyprus

[2] Avantama Ltd, Staefa, Laubisrutistr. 50, CH-8712, Switzerland

*Corresponding Author: Prof. Stelios A. Choulis
e-mail: stelios.choulis@cut.ac.cy

# Materials and Methods

***Materials*:**

Prepatterned glass-ITO substrates (sheet resistance $4\,\Omega\,\text{sq}^{-1}$) were purchased from Psiotec Ltd., P3HT from Rieke Metals, IDTBR from Solarmer, PC[60]BM from Solenne BV. PEI and all the other chemicals used in this study were purchased from Sigma-Aldrich. Whereas 10 at% antimony doped tin oxide (10 at% $Sb:SnO_2$) solution in mixture of butanols was developed from Avantama (ATO, Product-No. 10095). Details for Synthesis, properties and characterization of ATO can be found in the main text of the manuscript.

***ZnO sol gel synthesis*:** The ZnO films has been prepared using a zinc acetate dehydrate and monoethanolamine as a stabilizer dissolved in 2-methoxyethanol. 0.05 g of zinc acetate dehydrate has been mixed with 0.0142 g of monoethanolamine and dissolved in 0.5 ml of 2-methoxyethanol. The resulting precursor solution has being stirred at room temperature for 20 min in ambient conditions and then has been coated, using doctor blading, on top of ITO substrates and annealed after deposition, at 140 ºC for 20 min in air, forming a 40 nm thick ZnO layer.

***Device Fabrication*:**

The inverted OPVs under study was ITO/ETL/P3HT:PCBM/$MoO_3$/Ag. ITO substrates were sonicated in acetone and subsequently in isopropanol for 10 min. The ATO solution was deposited by doctor blade at 70 °C and annealed for 10 minutes at 120 °C. The PEI interfacial layer was spin coated resulting in an ultrathin layer and annealed at 100 °C for 10 minutes. Since the PTE interfacial layer in the range of 5 nm, the PEI processing step can be also considered as interface conditioning step of ATO. The PEI The active layer solution P3HT:PCBM was deposited on the ETL by doctor blade resulting in a film with a thickness of ~200 nm and annealed for 20 minutes at 140 °C prior to evaporation. Finally, 10 nm of $MoO_3$ and 100 nm Ag layers were thermally evaporated through a shadow mask to finalize the devices giving an active area of 0.9 mm$^2$. Encapsulation was applied directly after evaporation in the glove box using a glass coverslip and an Ossila E131 encapsulation epoxy resin activated by 365 nm UV irradiation.

*Characterization:*

The thickness of the films was measured with a Veeco Dektak 150 profilometer. AFM images were obtained using a Nanosurf easy scan 2 controller under the tapping mode. Electrical conductivity measurements were performed using a four-point microposition probe, Jandel MODEL RM3000. For illuminated current density–voltage (*J/V*) characteristics, a calibrated Newport Solar simulator equipped with a Xe lamp was used, providing an AM1.5G spectrum at 100 mW cm$^{-2}$ as measured by a certified oriel 91150 V calibration cell. Wide-angle XRD patterns were collected on a PANanalytical X´pert Pro MPD X-ray diffractometer equipped with a Cu (*λ*=1.5418 Å) rotating anode operated at 40 mA and 45 kV, in the Bragg-Brentano geometry. Transmittance and absorption measurements were performed with a Schimadzu UV-2700 UV-Vis optical spectrophotometer.